\documentclass{article}
\usepackage[margin=0.9in]{geometry}
\usepackage[version=3]{mhchem} 
\usepackage{siunitx} 
\usepackage{graphicx} 
\usepackage{amsmath} 
\usepackage{subfigure}
\usepackage{graphicx, subfigure}			
\usepackage{dcolumn}			
\usepackage{bm}					
\usepackage{float}			
\usepackage{amsmath,amsfonts}	
\usepackage{url}					
\usepackage{setspace}		
\usepackage{lineno}   			
\usepackage{multirow}

\title{\textbf{Exploring high-frequency eddy-current testing for sub-aperture defect characterisation using parametric-manifold mapping}} 

\author{Robert Hughes and Bruce W. Drinkwater} 

\date{} 

\begin{document}
\title{Exploring high-frequency eddy-current testing for sub-aperture defect characterisation using parametric-manifold mapping}
%
%

\author{Robert R. Hughes and Bruce W. Drinkwater} 

%

\maketitle              

\begin{center}
Department of Mechanical Engineering, University of Bristol,\\
Queens Building, University Walk, Bristol BS8 1TR, UK
\end{center}

\begin{abstract}
Accurate characterisation of small defects remains a challenge in non-destructive testing (NDT). 
In this paper, a principle-component parametric-manifold mapping approach is applied to single-frequency eddy-current defect characterisation problems for surface breaking defects in a planar half-space. 
A broad 1-8 MHz frequency-range FE-circuit model \& calibration approach is developed \& validated to simulate eddy-current scans of surface-breaking notch defects.  This model is used to generate parametric defect databases for surface breaking defects in an aluminium planar half-space and defect characterisation of experimental measurements performed. 
Parametric-manifold mapping was conducted in N-dimensional principle component space, reducing the dimensionality of the characterisation problem. In a study characterising slot depth, the model \& characterisation approach is shown to accurately invert the depth with greater accuracy than a simple amplitude inversion method with normalised percentage characterisation errors of $\pm 38 \%$ and $\pm 17 \%$ respectively measured at 2.0 MHz across 5 slot depths between 0.26 - 2.15 mm.  The approach is used to characterise the depth of a sloped slot demonstrating good accuracy up to $\approx 2.0 mm$ in depth over a broad range of sub-resonance frequencies, indicating applications in geometric feature inversion. Finally the technique is applied to finite rectangular notch defects of surface extents smaller than the diameter of the inspection coil (sub-aperture) over a range of frequencies. The results highlight the limitations in characterising these defects and indicate how the inherent instabilities in resonance can severely limit characterisation at these frequencies.
\end{abstract}


\section{Introduction}

Non-destructive evaluation (NDE) methods are used to detect, size and characterise defects that pose a threat to the structural integrity of safety, and operation critical industrial components \cite{Bray1992}. Unchecked defects can cause the failure of important infrastructure (transport, aerospace, oil and power). Accurate defect characterisation can provide valuable information about the physical nature and potential cause of a material defect and so is extremely useful for predicting the remaining lifespan of a component. Achieving accurate characterisation therefore has the potential to reduce the number of unnecessary and costly replacements of components.

The problem of sizing crack-like defects has been studied extensively in the literature, and various sizing inversion techniques have been proposed for eddy-current inspection analysis in a range of applications \cite{Sabbagh2013, Ahmed2017, Aldrin2014, Cherry2014}. Standard iterative approaches generate simulated trial solutions from a validated forward models to minimise a suitable cost function between experimental and simulated measurements \cite{Salucci2017, Rocca2009}. Iterative inversion methods are inherently computationally expensive and time-consuming, making real-time inversion difficult. As such non-iterative approaches, employing databases of off-line generated defect responses, have become popular, promising real-time characterisation \cite{Ahmed2019}.  Non-iterative approaches compute a database of possible solutions off-line for a given inspection geometry and defect parameter space, then use various techniques and algorithms to search and interpolate the database to find the closest matching results to experimental measurements. Authors have approached this from various directions in eddy-current including methods based in; machine learning techniques \cite{Miorelli2019, Rosado2013, Buck2016, Shokralla2016, DAngelo2018} and meta-model generation \cite{Cai2018, Douvenot2011, Salucci2019}. In spite of the impressive characterisation ability of these approaches to large defects, there remains a fundamental limit to the ability of these techniques to accurately size small surface-breaking defects, where sizing is of critical importance.  This limit is, in theory, determined by the dimensions of the inspection coils used, and manifests itself as densely populated regions in the lower-dimensional parameter space representing the defect dimensions.  The meaning of this is that any experimental, or realistic simulated result, that finds itself in this region will have many potential matches from the parameter space, increasing the characterisation uncertainty.  Most work to date has concentrated on low-frequency ($< 1 MHz$) inspections of defects whose surface extent is larger than the coil diameter (referred to in this paper as the aperture). However, most small-defect ECT inspections regularly employ higher frequencies to reduce the depth of penetration and increases sensitivity. In this paper we aim to explore the modelling and defect characterisation potential of high-frequency ($>1 MHz$) ECT measurements, and to examine whether any frequency dependence exists in the characterisation accuracy of surface breaking defects in aluminium.

In this paper, we will not focus on the efficiency of different methods of sampling the parameter space or optimising characterisation algorithms, as this is the subject of many ongoing studies by other authors as highlighted above.  Instead, we will employ a similar technique, the non-iterative parametric-manifold mapping method, developed for ultrasonic array defect characterisation by Velichko et. al. \cite{Velichko2017}, to characterise sub-aperture defects in a semi-infinite aluminium (Al) half-space, from high-frequency eddy-current measurements. 

This was approached by developing a bespoke hybrid finite-element (FE)/circuit model to simulate the behaviour of eddy-current sensors in the presence of notch-like defects. This model was then optimised and validated against experimental measurements. A simulated defect database was then generated, principle-component (pc) analysis performed and the database converted into a reduced dimensional manifold in pc-space. Each point on the manifold represents the expected ECT response of a given defect size in the lower dimensional pc-space. Experimental sub-aperture defect scans were conducted over a broad frequency range and characterised by evaluating their euclidean distance to the manifold in pc-space.



\section{Methodology}
The characterisation of surface-breaking defects was performed on 1\&2D eddy-current surface scans in Al samples. The methodology developed for ultrasonic phased-array defect characterisation utilises the parametric manifold approach to find the closest matching simulated defect scattering-matrix to any given experimentally measured defect scattering-matrix within a pre-simulated database \cite{Velichko2017}. In this way, the size of an experimentally measured defect can be characterised. In this paper, the impedance scan data at a single-frequency from ECT inspections is used as an analogue for the ultrasonic array scattering matrix used by Velichko et.al.\cite{Velichko2017}.  Instead of using the ultrasonic scattering matrices of defects, direct ECT scan data of slot \& notch defects in Al are used.

For each type of inspection (1D line scans of slots \& 2D area scans of notch defects), simulated databases of defect scans are produced from a bespoke hybrid FE-circuit model, and converted into a parametric manifold in lower-dimensional principle-component space (pc-space). Experimental inspections of defects are then converted into pc-space and characterised based on their Euclidean distance from the parametric manifold. 

In this paper, the methodology is divided into a number of sections detailing; the experimental measurement system, the optimisation and calibration of the bespoke hybrid-model, and the characterisation using the calibrated model.

\begin{figure}[!t]
\centering\includegraphics[width=6.5in]{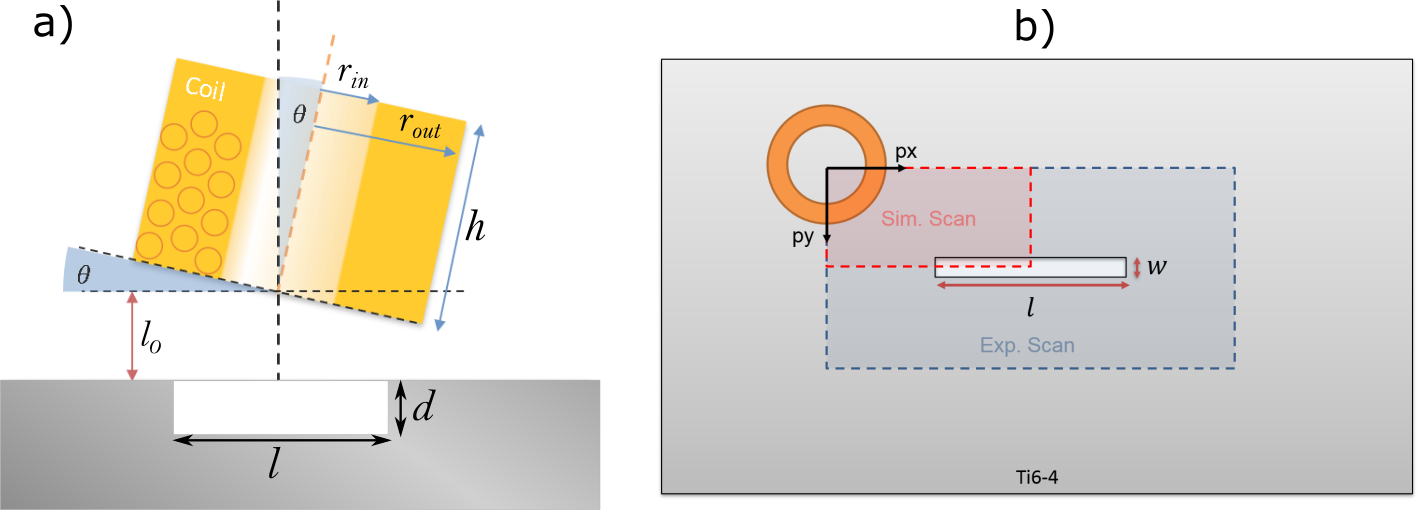}
\caption{Coil inspection geometry showing a) a cross-sectional diagram of the critical coil and defect dimensions, and b) a top-down diagram of the inspection setup identifying the experimental and simulated scan areas relative to the defect. }
\label{fig_scan}
\end{figure}

\subsection{Experimental Setup}\label{sec_expset}
This study required the development of a bespoke ECT model for a given sensor coil design.  Experimental measurements were recorded to optimise and calibrate the model prior to the development of a defect database.  Details of the experimental methods are given below.

\subsubsection{Coil Design}\label{sec_coil}
A $1.5$ mm high, air-cored, $3.0$ mm outer-diameter (OD), solenoid coil depicted in Figure \ref{fig_scan}.a was used as the induction sensor throughout the study. Solenoid coil sensors are commonly used in ECT measurements. The dimensions were selected to simplify the finite-element (FE) modelling of the sensor, as smaller coil dimensions become difficult to model accurately due to deviations from the assumption of uniform current distributions of the coil turns. The shape and critical dimensions of the coil are shown in Table \ref{tab_coil}. The coil had a measured free-space inductance of $9.66 \pm 0.01 \mu H$.  The sensor coil is embedded in a plastic probe tip and secured with glue. This glue also forms a thin layer protecting the coil from wear during inspection (see step 3 in Figure~\ref{fig_cal_process}) and must be calibrate out during the model development process.

The model developed in this study makes a number of assumptions for computational simplicity: Coil tilt, $\theta$, is negligible ($\approx 0^o$). Lift-off from the material surface, $h$, is constant throughout the defect scan. Defects are rectangular voids, and
electrostatic interactions are not considered.

\vspace*{-5pt}

\begin{table}[!h]
\caption{Measured coil sensor properties}
\label{tab_coil}
\centering\begin{tabular}{lll}
\hline
Physical Property & Value & Units \\
\hline
$h$ & $1.5\pm 0.1$ & mm \\
$d_{in}$ & $1.75 \pm 0.05$ & mm \\
$d_{out}$ & $3.0\pm 0.1$ & mm \\
$w_t$ & $0.10\pm 0.01$ & mm\\
$N$ & $75\pm 1$ & dim. \\
$n_L$ & $5\pm 1$ & dim. \\
\hline
Electrical Property & Value & Units \\
\hline
$L_0$ & $9.66\pm 0.01$ & $\mu H$ \\
\hline
\end{tabular}
\vspace*{-4pt}
\end{table}

\subsubsection{Impedance Spectra Measurement}\label{sec_expMeth}
The coil sensor defined in section \ref{sec_coil} was mounted into a purpose built ECT probe holder, providing continuous surface normal pressure to the coil (via an axial compression spring) keeping it flush to the inspection surface during scans. This minimised the risk of potential standoff variations over the course of the inspection, whilst maintaining the coil normal to the surface $\pm 1^o$.

The coil was wired into the probe using $100 \pm 5 mm$ Dupont wires and connected to an impedance analyser (TE3001 TREWMAC, Scott Creek, Australia)  via a $500 \pm 5 mm$ long shielded twisted pair cable  (8451 multi-conductor - single pair cable from Belden, St. Louis, MO, US). The shielded twisted pair was used to minimise the overall cable capacitance and potential noise due to cable movement during scans. 

\begin{figure}[!b]
\centering\includegraphics[width=6in]{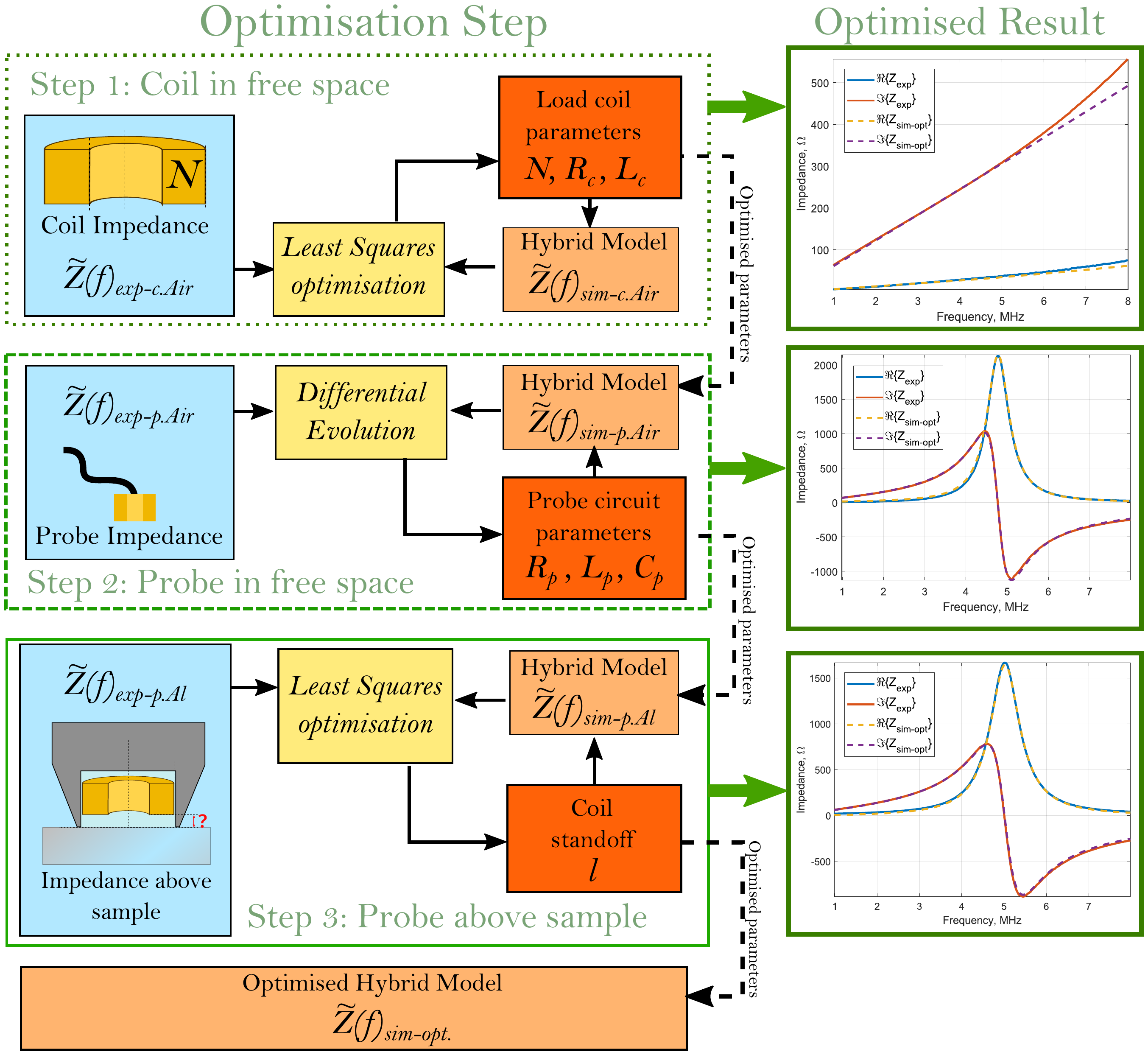}
\caption{Flow diagram of the model optimisation method, showing the stages of optimisation. Step 1: Optimisation of FE model parameters to match experimentally measured coil inductance. Step 2: Differential evolution optimisation of probe (coil and twisted pair in parallel) equivalent circuit variables to match experimentally measured impedance. Step 3: Least square optimisation of material standoff parameter in FE simulation.}
\label{fig_block}
\end{figure}

The probe is attached to a high-resolution Zaber (Vancouver, British Columbia, Canada) X-LSM150A XY scanning stage, used to perform raster scans of specimens. Each scan has a spatial resolution of $\Delta x = \Delta y = 0.2$mm. A bespoke MATLAB control program was developed to perform a point-by-point raster scan inspection, at each point a full 100 point complex impedance frequency spectra, $\tilde{Z}(f)_{exp}$, between 1-8 MHz is recorded with the TE3001 impedance analyser.

\subsection{Modelling Coil Impedance}\label{sec_sim}
A 3-dimensional COMSOL 5.3a model of the inductive coil senor was developed using the AC/DC module and the coil parameters specified in section~\ref{sec_coil} to generate a multi-turn coil model.  This model was then extended to include a planar test specimen beneath the coil at stand-off $h$ and with tilt $\theta$ (see Figure~\ref{fig_scan}.a) relative to the normal axis of the coil.  In order to calibrate out known unknown parameters from the measurements, the FE impedance model was combined with an equivalent circuit model of the inspection probe. Any variability of parameters in the model to expected values can then be optimised out a fixed to provide a more accurate comparison between the simulated and experimental inspection data.

\begin{figure}[!h]
\centering\includegraphics[width=4in]{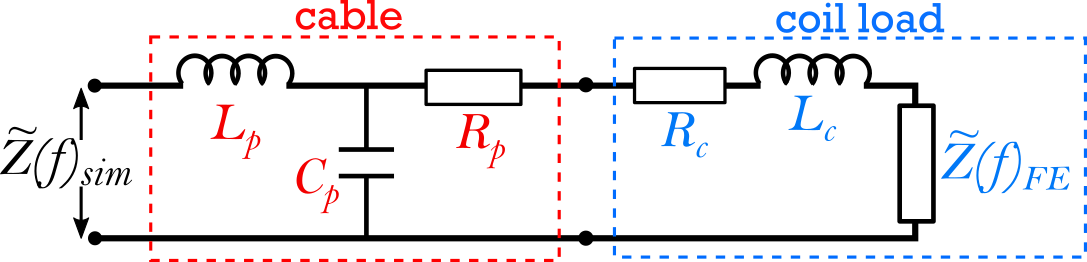}
\caption{Hyrbid model equivalence circuit diagram}
\label{fig_circ}
\end{figure}

In order to validate and relate the changes in measured coil impedance to the experimentally measured impedance, the FE-simulated coil impedance frequency spectra, $\tilde{Z}(f)_{FE}$, is input into a simple equivalence circuit model as shown in Figure~\ref{fig_circ}, and evaluated using the equation,

\begin{equation}
    \tilde{Z}(f)_{sim} = \tilde{Z}(f)_C \frac{R_p + \tilde{Z}(f)_L}{R_p + \tilde{Z}(f)_L + \tilde{Z}(f)_C} + i\omega L_p ,
\end{equation}
where,
\begin{align}
    \tilde{Z}(f)_L &= R_c + i \omega L_c + \tilde{Z}(f)_{FE},\\
    \tilde{Z}(f)_C &= \frac{1}{i\omega C_p}.
\end{align}

Here, the capacitance, resistance and inductance of the connecting twisted-pair cable are defined as $C_p$, $R_p$, and $L_p$ respectively. The resulting FE-Circuit simulated impedance, $\tilde{Z}(f)_{sim}$, produces the simulated impedance spectra of the whole inspection probe. 

Before this hybrid model is used to predict defect signals, the FE model parameters and unknown circuit variables must first be optimised to accurately represent the probe in free-space (air). Secondly the (immeasurable) inherent inspection stand-off $h$ of the probe must be optimised within the FE model. This is achieved by performing a series of optimisation steps shown in Figure~\ref{fig_block} and detailed below.  The optimised model was then used to perform a parametric study for a range of rectangular defect dimensions; surface length, $1.0<l<7.0 mm$, and notch depth $0.1<d<3.0 mm$.

\subsubsection{FE Model Optimisation: Air}
An experimental impedance measurement of the inspection coil in air, $\tilde{Z}(f)_{exp-c.Air}$, is made and used to optimise the ill-defined parameter of the number of coil turns, $N$, in the FE model in free-space. Although the number of turns was counted during the manufacture of the coil, there is an uncertainty of $\pm 1$ turn due to manufacturing challenges. To ensure the model is as accurate as possible, the number of turns is checked and optimised to ensure that the overall inductance of the coil matches experimental measurements.  A systematic integer increase in $N$ is performed in the FE-model to identify the optimum value for $N$, which gives the closest match to the experimental inductance, $L$, as defined by the gradient of the imaginary component of impedance, $\nabla \Im \{Z\}$, within the linear regime of the experimental curve (i.e. $1-4$ MHz). An additional frequency-dependant resistance component, $R_c(\omega)$, and inductive tuning component $L_c$, must also be added to the FE modelled coil impedance to match the experimental result.  Together, $\tilde{Z}(f)_{FE-Air}$, $R(\omega)_c$ and $L_c$ in series equal the simulated coil load impedance, $\tilde{Z}(f)_L$ (see Figure~\ref{fig_circ}). The impedance and circuit components defining the coil in air are referenced symbolically with the sub-script letter $c.Air$.


\subsubsection{Circuit Model Optimisation: Air}
An experimental impedance measurement is made of the coil connected to the measurement instrument via a $500 \pm 5mm$ twisted-pair cable. 
This is referred to as the probe impedance, $\tilde{Z}_{exp-p.Air}$, and is differentiated symbolically from coil variables and components by the sub-script letter $p$. 
The optimised simulated coil impedance in air, $\tilde{Z}(f)_L$, is input into the circuit model shown in Figure~\ref{fig_circ} to simulate the impedance spectrum of the probe (coil and cable) in air. The measured experimental impedance is used to find optimum parameters for the unknown components of the equivalent circuit in the hybrid model shown in Figure~\ref{fig_circ}, implementing an open-source differential evolution optimisation MATLAB program (copyright 2017, Markus Buehren, \cite{MarkusBuehren}) to optimise the unknown circuit parameters ($R_p$, $L_p$ and $C_p$) based on least-squares minimisation between the experimental and modelled impedance spectra. 

\begin{table}[!h]
\caption{Optimised circuit parameters}
\label{tab_stg2_params}
\centering\begin{tabular}{llll}
\hline
Variable & Optimised Value & Units \\
\hline
$R_p$ & $4.6 \pm 0.1$ & $\Omega$ \\
$L_p$ & $0.7 \pm 0.1$ & $\mu H$ \\
$C_p$ & $113 \pm 1$ & $pF$ \\
$R_c$ & $\omega (1.3\times10^{-6}) - 6$ & $\Omega$ \\
$L_c$ & $0.05 \pm 0.01$ & $uH$ \\
\hline
\end{tabular}
\vspace*{-4pt}
\end{table}

The optimised values defined in Table~\ref{tab_stg2_params} were used to produce the optimum simulated impedance profile for the probe in air shown in Figure~\ref{fig_block} compared to the experimentally measured impedance profile.

\subsubsection{Coil Stand-off Optimisation}
The third stage of model optimisation uses experimentally measured impedance profiles of the coil above test materials, to determine the inherent stand-off of the coil from the surface of any given inspection material. This is difficult to measure feature of the experimental set-up of the probe (see section~\ref{sec_coil}). The thickness of this layer must be evaluated in order to produce a valid model of the sensor set-up, and is evaluated via inversion of an experimental measurement of the sensor above the surface of a material, as shown in Figure~\ref{fig_block}. In future iterations of the procedure, the coil tilt could also be calibrated in the same way. This procedure was performed for measurements made above Al and Ti6-4 to test the validity of the model.
The result of the stand-off inversion demonstrated that the model predicts an inherent stand-off from the material of $l_0 = 0.30\pm 0.03 mm$ for both Al and Ti6-4. This parameter was fixed in the FE-model and stand-off inversion performed to validate the model and confirm the expected stand-off for defect scans (i.e. coil protected by Kapton tape layer).

\subsubsection{Model Validation: Stand-off inversion}
To validate the model, experimental impedance spectra measurements were made of the sensor at well-defined stand-offs from the surface of Al and Ti6-4. The FE-model was then used to iterate through stand-off distances to generate $\tilde{Z}(f)_{FE}$ predictions and least-squares linear interpolation used to estimate the experimental stand-off. The model was tested in this way at 3 stand-off distances using physical dielectric spacers; a 1.0 mm glass microscope slide, a 1.5 mm glass slide and Kapton tape (typical inspection stand-off). Each spacer was measured experimentally with a digital micrometer in five locations and the average taken.


The measured and model-estimated distances of these stand-offs are shown in Table~\ref{tab_standoffVal}.  The resulting estimations demonstrate the model is valid over the whole stand-off range above Al but discrepancies arise at higher stand-offs in the Ti6-4 model. The reason for this is not clear, but could be due to Ti6-4's comparatively low conductivity and may result from unaccounted for electrostatic interactions between the coil and the surface that are otherwise negligible compared to the inductive interactions when above a more highly conducting material (Al). However, this process serves to validate the model for typical inspection stand-offs such that it can be used for the simulation of defect scans.

\begin{table}[!h]
\caption{Stand-off inversion}
\label{tab_standoffVal}
\centering\begin{tabular}{lccc}
\hline
Stand-off spacer & Exp. $\pm 0.02$ [mm] & Est. above Al $\pm 0.03$ [mm] & Est. above Ti6-4 $\pm 0.03$ [mm]\\
\hline
Thick glass slide & $1.03$ & $1.07$ & $0.77$ \\
Thin glass slide & $0.16$ & $0.15$ & $0.14$ \\
Kapton tape & $0.06$ & $0.07$ & $0.07$ \\\hline
\end{tabular}
\vspace*{-4pt}
\end{table}

This concludes the optimisation of the hybrid FE-Circuit model developed. Below is the list of optimised FE-model parameters. At this stage, each of the parameters in Tables~\ref{tab_stg3_params} and~\ref{tab_stg2_params} are fixed in the hybrid-model and databases of defect scans can be produced.

\begin{table}[!h]
\caption{Optimised FE-model parameters}
\label{tab_stg3_params}
\centering\begin{tabular}{lcl}
\hline
COMSOL Property & Value & Units \\
\hline
$N$ & $76$ & $turns$ \\
$l_{0}$ & $0.30 \pm 0.03$ & $mm$ \\
$l_{tape}$ & $0.07 \pm 0.03$ & $mm$ \\\hline
\end{tabular}
\vspace*{-4pt}
\end{table}

\subsection{Characterisation using Calibrated Model}
The model detailed above was used to generate a parametric database of ECT scans of defects with varying dimensions in different materials. ECT scans are simulated by calculating the complex impedance spectra, $\tilde{Z}_{sim}(f)$, at incremental locations relative to the central defect axis defined by the spatial coordinate vector $\mathbf{r}$. Each ECT defect scan can then be defined by what we will refer to as its impedance matrix or \textit{Z-matrix} $\tilde{Z}_{sim}(\mathbf{r},f)$, where $f$ is the frequency axes of size $P$. For 1D line-scans over infinite length slots $\mathbf{r}= x$, and 2D surface scans over finite length notches $\mathbf{r}= \{x,y\}$, where $x$ and $y$ are the planar orthogonal coordinate vectors of the scan above the specimen with size $N_x \times N_y$ respectively. As such a 2D defect scan has a Z-matrix, $\tilde{Z}_{sim}(x,y,f)$, of size $N_x \times N_y \times N_f$.

Z-matrices can be simulated for a series of parametrically varying defects to produce a database of Z-matrices (simulated scans) for use in the manifold-mapping characterisation approach.  However, before characterisation can be performed, the simulated defect Z-matrices need to be calibrated to a defect of known dimensions. This process calibrates out any unmodelled behaviour, such as probe drag, differences between simulated and experimental conductivity or discrepancies between manufactured calibration defect dimensions and those simulated.

\subsubsection{Defect Calibration} \label{sec_cal}
\begin{figure}[!h]
\centering\includegraphics[width=6.5in]{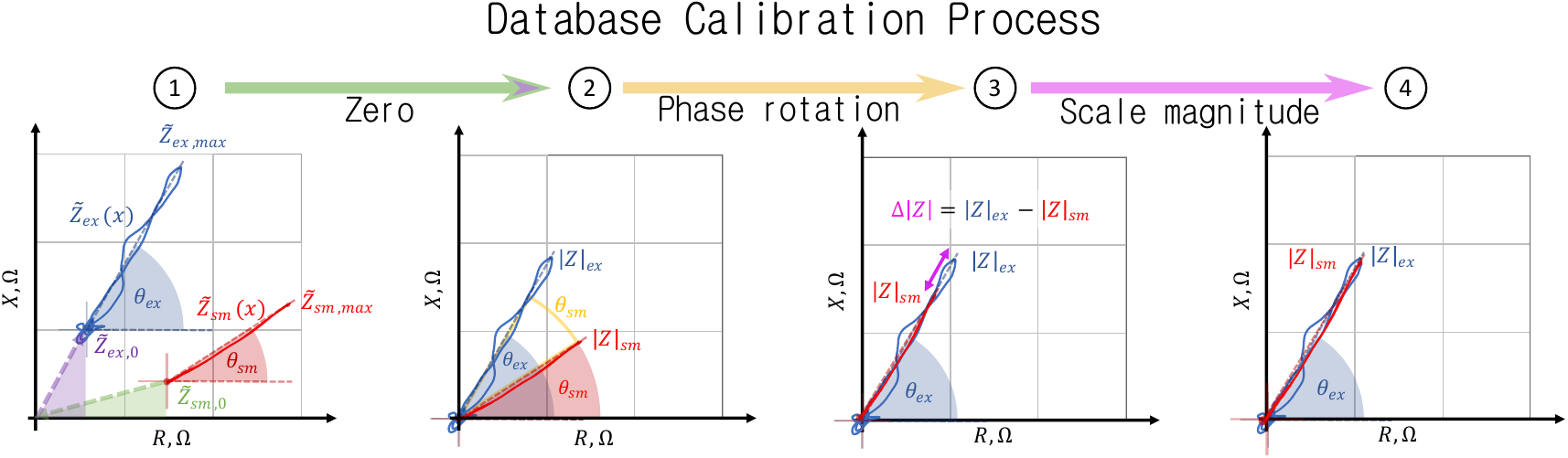}
\caption{Database calibration process, showing the complex space transforms applied to simulated data $\tilde{Z}_{sm}(x)$ with respect to an experimentally measured calibration slot $\tilde{Z}_{ex}(x)$.}
\label{fig_cal_process}
\end{figure}

The calibration process (as demonstrated for a single frequency, $f$, in Figure~\ref{fig_cal_process}) involves simulating the Z-matrix, $Z_{sim} = \tilde{Z}_{sim}(\mathbf{r},f)$, and comparing this to the experimentally recorded defect scan, $Z_{exp} = \tilde{Z}_{exp}(\mathbf{r},f)$, for a defect of known dimensions. Complex space (Argand) transforms (steps 1-4) are applied to $Z_{sim}$ to calibrate the phase angle and magnitude to $Z_{exp}$. These complex-space transforms are then applied to all simulated scans to calibrate them ready for parametric-manifold mapping characterisation.  

Figure~\ref{fig_cal_process} shows the calibration process for the Z-matrix at a single frequency, but the process is applied for the whole frequency spectrum.  Each Z-matrix can be represented in complex-space by a peak vector about the origin (step 2) $\mathbf{Z}(f) = \mathbf{Z_{max}}(f) - \mathbf{Z_0}(f)$, where $\mathbf{Z_0}(f)=\{R_0(f), X_0(f)\}$ and $\mathbf{Z_{max}}(f)=\{R_{max}(f), X_{max}(f)\}$. $\mathbf{Z_0}(f)$ and $\mathbf{Z_{max}}(f)$ are the mean impedance spectra of the sensor over undamaged material (background) and the peak impedance spectra of the calibration defect response respectively. The disparity in the phase angle, $\theta_{\Delta}(f)$, between the experimental and simulated vectors is applied as a rotation about the origin to the simulated data (step 3). The disparity in vector magnitudes $\Delta|Z|$ is applied as a scaling to the simulated data.  The calibration steps applied to all simulated data are detailed below: 
\begin{itemize}
    \item \textbf{Zero} - Background subtraction (or zeroing) each Z-matrix in complex-space by subtracting $Z_0(f)$ from $Z = \tilde{Z}(\mathbf{r},f)$ of both simulated and experimental.
    \item \textbf{Phase rotation} - Applying a phase rotation, $\theta_\Delta$, in complex space to align the simulated indication with the experimental calibration signal.
    \item \textbf{Scale magnitude} - Finally a scaling factor is applied to force the simulated peak magnitude to match the experimental peak magnitude.
\end{itemize}

\subsubsection{Principal-Component (PC) Defect Manifold}\label{sec_pca}
Principal-component analysis (PCA) is used to reduce the defect scan database down to a handful of dominant metrics which can be used to identify each defect scan. The largest N pc-components of these can be plotted against one another to produce a manifold ND PC-space. Experimental measurements can be transformed into the same ND PC-space and their proximity to points on the manifold used to characterise defect dimensions (see Velichko et. al. \cite{Velichko2017} for details on the manifold generation process). 

We represent the complex impedance spectra scan data measured by the inspection system (see section~\ref{sec_expset}) as a matrix, 
$Z=\tilde{Z}(x,y,f)$, where coordinates $[x,y]$ define the spatial scan region and $f$ denotes the frequency spectrum. We have defined $Z$ as the complex impedance matrix or \textit{Z-matrix} of size $[N_x, N_y, N_f]$ corresponding to the sizes of each matrix dimension $x$, $y$ and $f$.  Each Z-matrix, $Z$, can be represented as a vector $\mathbf{z}={z1,...,z_{N_z}}^T$, where the $i^{th}$ term $z_i = \tilde{Z}(x_n,y_m,f_l)$ in an $N_z = N_x \times N_y \times N_f$-dimensional space. A parametric database of Z-matrices for a range of defects can therefore be defined as, $Z_p = \tilde{Z}(x,y,f; \mathbf{p})$, where $\mathbf{p} = \{p_1,...,p_{Np}\}$ represents a set of $N_p$ parameters defining the defect class (i.e. $N_p=2$ for EDM notches parameterised by length and depth). A complete set of Z-matrices in a given defect class can therefore be represented by the $N_p \times M$ matrix
\begin{equation}
    \mathbf{Z_p}=\{\mathbf{z_p}(\mathbf{p_1}),..., \mathbf{z_p}(\mathbf{p_M})\}\in \mathbb{R}^{N_z \times M},
\end{equation}\label{eqn_Zp}
where $M$ is the total number of Z-matrices in the defect class and $\mathbf{z_p}$ is the vector corresponding to the defect Z-matrix $Z_p$.

As shown in Velichko et.al 2017 \cite{Velichko2017}, principal component analysis can be used to obtain a reduced-dimensional coordinate system through the singular value decomposition of the covariance matrix of $\mathbf{Z_p}$,
\begin{equation}
    \mathbf{R} = \mathbf{V}\mathbf{D}\mathbf{V}^T,
\end{equation}\label{eqn_covar}
where $\mathbf{R}$ is the $N_z \times N_z$ covariance matrix of $\mathbf{Z_p}$, $\mathbf{D}$ is the diagonal matrix containing the eigenvalues of $\mathbf{R}$, and the axes of the new coordinate system is now given by the column vectors of $\mathbf{V}$.  As such any Z-matrix, $\mathbf{z}$, in the new pc coordinate system is represented by the vector,
\begin{equation}
    \mathbf{z}^{(pc)} = \mathbf{V}^T\mathbf{z}.
\end{equation}\label{eqn_spc}
Following PCA, the dominant features of the ECT response to the defect class is effectively contained within a lower-dimensional space, termed principal-component space (or pc-space), where the first $N_{pc}$ coordinates account for the majority of the variation in the Z-matrices between different defects. We can set $N_{pc}$ to be equal to the number of eigenvalues of the covariance matrix $\mathbf{R}$ above a given threshold.  For this study the eigenvalue threshold was defined as 0.01 (see Figure~\ref{fig_cal_slot_result}.b).

\section{Results \& Discussion}
The model was used to generate databases of simulated defect inspection measurements. In this investigation, 2 studies were considered: The simple infinitely long slot, and the finite-length notch performed in Al. In both cases, defect characterisation of experimental measurements using the pc-manifold approach was explored.  All characterisation is performed on single-frequency scan data. For this section 2.0 MHz was selected as the operation frequency.

\subsection{Slot Depth Characterisation}
\begin{figure}[!h]
\centering\includegraphics[width=6.5in]{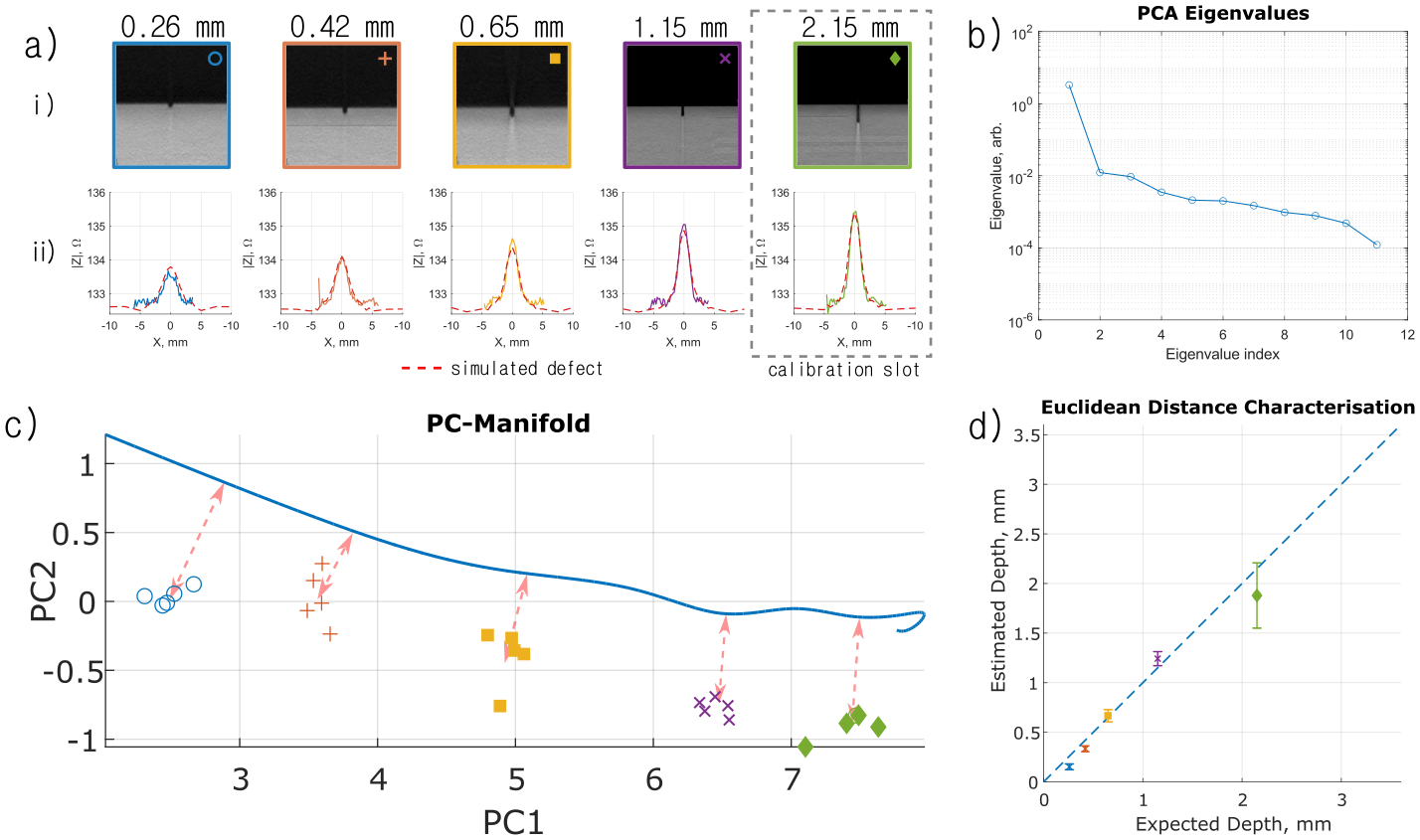}
\caption{Slot depth validation of calibration process (a) and inversion (b-d) of experimental measurements at 2.0 MHz. Showing a) the X-ray cross-section of 5 wire-cut slots (i) and the experimental and calibrated simulation scan data for equivalent slot depths (ii). Figure b) shows the principle component eigenvalues for the first 11 principle components of the simulated slot database. Figure c) shows the 2D pc-space manifold (blue line) relative to the experimental measurements of the wire-cut slots, highlighting (red dashed arrow) the euclidean distance from the mean experimental value to the simulated manifold in pc-space.  Figure d) shows the defect depth characterisation using the euclidean distance method.}
\label{fig_cal_slot_result}
\end{figure}

The optimised hybrid-model was used to generate a parametric database of ECT responses to wire-cut slots of varying depth in Al. One-dimensional line-scans, $\tilde{Z}(x,f)$, (see Figure~\ref{fig_cal_slot_result}) were simulated with scanning resolution of $\Delta x = 0.5\ mm$ between $0 \leq x \leq 10.0\ mm$ perpendicular to the axis of the slot, with the origin at the central axis of the slot. Slot depths were simulated in increments of $\Delta d = 0.3 mm$ between $0.1 \leq d \leq 3.1 mm$ with rectangular cross-sections. 

Figure~\ref{fig_cal_slot_result}.a shows the calibrated simulation results compared to experimental data obtained for a range of slot depths in aluminium, where the database has been calibrated to the experimental data for a $2.15 mm$ deep wire-cut slot (slot width $=0.3mm$)\footnote{All defect dimensions are measured with an error of $\pm 0.03 mm$ from X-ray CT data unless stated otherwise.}. From the calibrated simulated database of slot depth line-scans, PCA can be performed to reduce the dimensionality of the database.

Figure~\ref{fig_cal_slot_result}.b shows the eigenvalues of the first 11 PC's of the slot depth database for the real and imaginary components of the data.  It is clear from the figure that this characterisation problem can be reduced down to only a handful of PC's as the eigenvalues fall dramatically after the initial PC.  It is expected that this dominant PC is the peak amplitude of the scan.  Therefore, a 2D PC manifold-mapping approach is validated in comparison to a simple amplitude inversion method. This comparison will determine whether any benefit is gained from PC manifold-mapping for this application.


Experimentally measured ECT responses to wire-cut slots in Aluminium were characterised in 2D pc-space, shown in Figure~\ref{fig_cal_slot_result}.c, by plotting the experimental defect measurements in pc-space with the simulated defect manifold (presented in 2D pc-space as the blue line).  The slot depth can be calculated by measuring the point on the manifold that has the shorted euclidean distance (the red dashed arrows in Figure~\ref{fig_cal_slot_result}.c) to each measurement. It is this point on the manifold that becomes the prediction of the slot depth shown in Figure~\ref{fig_cal_slot_result}.d.  The numerical values of the inversion are shown in Table~\ref{tab_slotDepth} compared to the rudimentary amplitude inversion method.  Each slot estimate is based on the median characterisation value and has an error equal to 2 times the standard deviation ($2\sigma$) from 5 repeat experimental measurements.

\begin{table}[!h]
\caption{Estimated slot depth for measurements at 2.0 MHz}
\label{tab_slotDepth}
\centering\begin{tabular}{ccc}
\hline
Measured, $\pm 0.03$ mm & Amplitude Inversion., mm & 2D PC-Manifold Estimate, mm \\
\hline
$0.26$ & $0.06 \pm 0.02$ & $0.15 \pm 0.03$ \\
$0.42$ & $0.23 \pm 0.04$ & $0.33 \pm 0.03$ \\
$0.65$ & $0.63 \pm 0.06$ & $0.67 \pm 0.06$ \\
$1.15$ & $1.14 \pm 0.09$ & $1.24 \pm 0.07$ \\
$2.15$ & $1.8 \pm 0.1$ & $1.9 \pm 0.3$ \\\hline
\end{tabular}
\vspace*{-4pt}
\end{table}

\begin{figure}[!t]
\centering\includegraphics[width=6.5in]{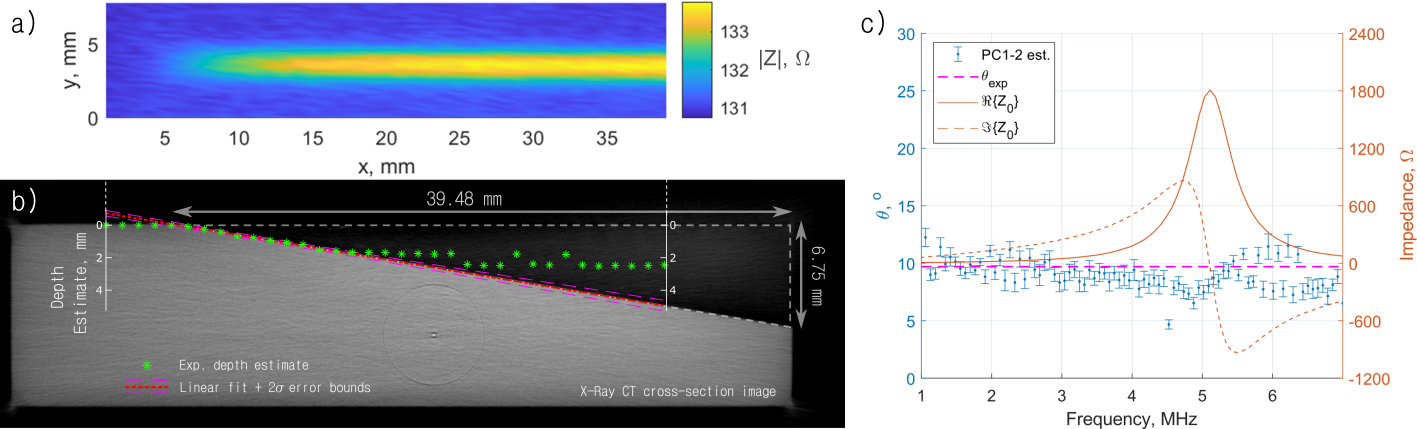}
\caption{Experimental slot-depth characterisation of 2.0 MHz ECT scans of a sloped slot in aluminium using the 2D pc-manifold approach, showing a) the raw ECT scan image, and b) the manifold-mapping estimated depth superimposed onto an X-ray CT cross-sectional image of the slot. A linear best fit is shown fitted to the linear portion ($5\le x \le 15 mm$) of the experimental depth estimate, along with $2\sigma$ error bounds. Graph c) shows the calculated slot-angle $\theta$, characterised by the linear fit, over the full frequency spectrum.  The predictions are shown relative to electrical resonance of the ECT probe on aluminium. Error bars represent 2 times the standard deviation.}
\label{fig_slope_slot}
\end{figure}

Although peak amplitude inversion is the most intuitive approach for characterising the experimental slot depths, the results in Table~\ref{tab_slotDepth} demonstrate how the 2D pc-manifold-mapping approach achieves a greater overall accuracy, with a normalised percentage error (NPE) across all slot depths of $\pm 17 \%$ relative to the X-ray CT measured values compared to the amplitude inversion method (NPE = $\pm 38 \%$). To further validate this method, the pc-manifold approach was applied to experimental measurements made along a sloped slot in Al (see Figure~\ref{fig_slope_slot}).

\subsubsection{Sloped Slot Characterisation}
From the depth characterisation graph shown in Figure~\ref{fig_slope_slot}, the angle of the sloped slot can be calculated and compared to the X-Ray CT measured angle.  The measured angle of the slope was $9.70 \pm 0.04 ^o$, and the ECT characterised angle at 2.0 MHz was $9.4 \pm 0.6 ^o$.  The results of the slot characterisation demonstrate that, for the experimental setup used, accurate inversion is limited to defects less than $2.0 mm$ in depth.  This is most likely limited by the geometric depth of penetration of the coil dimensions \cite{Smith2004}, and is therefore an expected limitation of the measurement.  Although this is a fundamental limitation of the technique, most ECT inspections are concerned with detecting and characterising defects shallower than $2 mm$ in depth, so the approach remains applicable. The approach was applied to all frequencies within the measured and simulated range to assess the accuracy of the model at all frequencies up to and beyond the electrical resonance of the probe.

\subsubsection{Frequency Evaluation}
The validity of the model was evaluated over the full measured frequency spectrum of the probe, by characterising the slot slope angle, $\theta$, over all measured frequencies, using the parametric manifold mapping approach in 2D pc-space (PC1-2).  Figure~\ref{fig_slope_slot}.c shows the experimentally characterised slot angle as a function of frequency, compared to the experimentally measured impedance spectra. The slot angle, $\theta(f)$, is calculated from linear best fits of the depth estimates between $5 \le x \le 15 mm$ (see Figure~\ref{fig_slope_slot}.b).  The results of this analysis is shown in Figure~\ref{fig_slope_slot}.c with reference to the X-ray CT measured angle, $\theta_{exp}$ and the impedance spectrum of the probe above aluminium, $\tilde{Z_0}$. 


The results in Figure~\ref{fig_slope_slot}.c demonstrate that the model can predict the impedance changes for frequencies below electrical resonance.  However, the model is shown to be inaccurate and inconsistent within a range of frequencies approaching and beyond electrical resonance (4.5-7.0 MHz).  This feature can be attributed to the instability of electrical resonance as the balance point between inductive and capactive elements within the measurement probe. Slight variations in inductive and capacitive components of the whole probe can cause incoherent shifting in the resonant frequency, which are difficult to model. 
In the final section, the manifold mapping approach is applied to simulated finite notch characterisation in aluminium. 

\subsection{Finite Notch Size Characterisation}
The hybrid model (see section~\ref{sec_sim}) was used to simulate a database of 2D ECT scans, $\tilde{Z}(x,y,f)$, of rectangular EDM notches in Al. Rectangular notches are typically characterised by 3 parameters - length, depth \& width. In this instance the width was kept constant wile the other defect parameters were varied. 

\begin{figure}[!h]
\centering\includegraphics[width=3.5in]{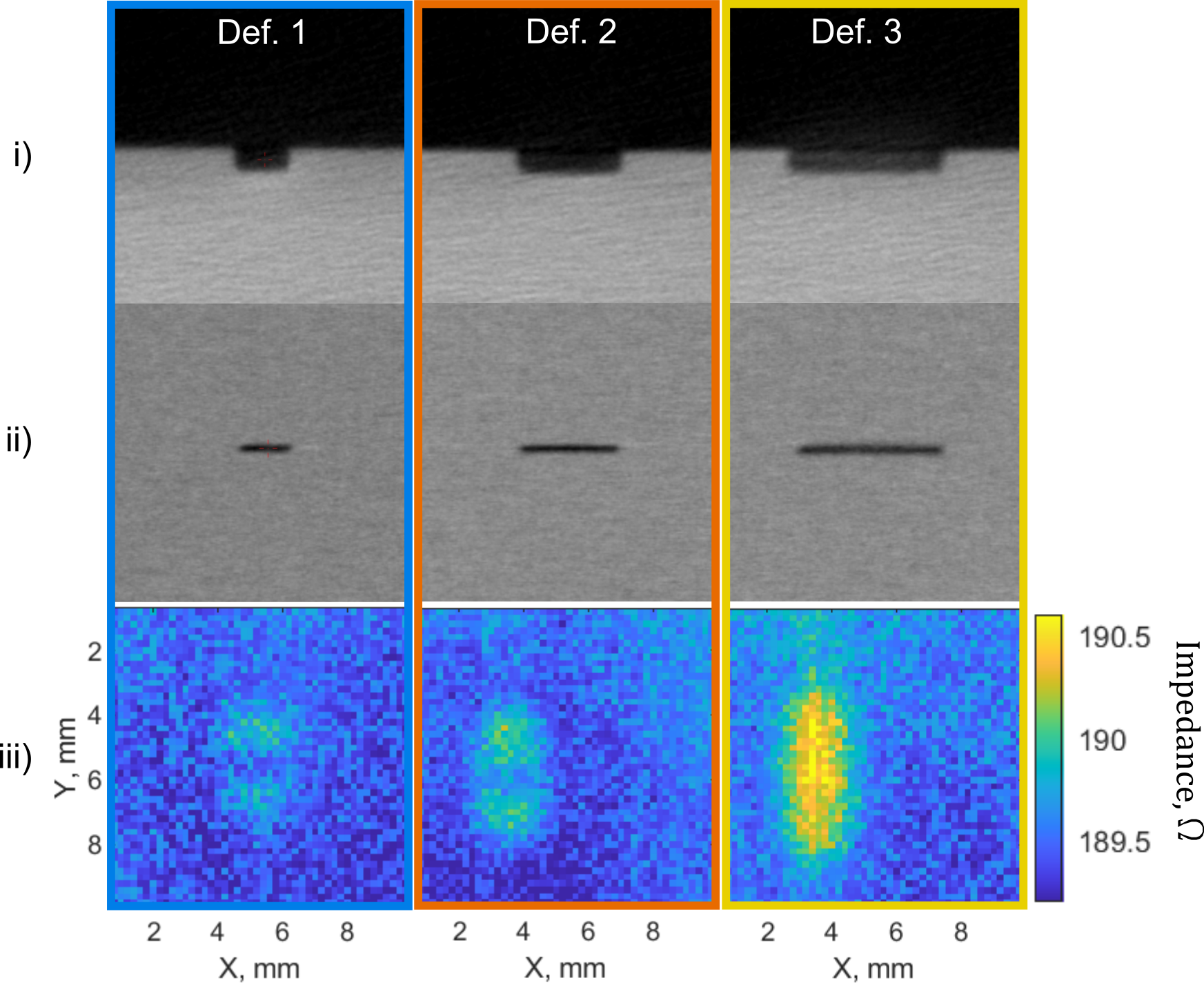}\includegraphics[width=3in]{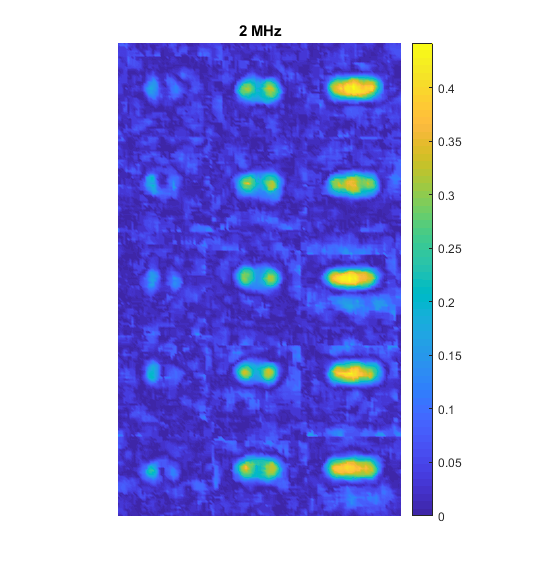}
\caption{Left: X-Ray computed tomography images of electro-discharge machined notch defects in Aluminium of the 3 trial defects, showing the size-on (i) and top-down (ii) projections of the defects compared to example raw eddy-current inspection image of the defects at 2MHz. Right: Filtered repeat defect scan of each defect at 2MHz.}
\label{fig_notch-CT}
\end{figure}

Three EDM notches of lengths smaller than the diameter of the inspection coil (sub-aperture), were manufactured in Al and their dimensions measured using X-ray CT imaging shown in Figure~\ref{fig_notch-CT}.i-ii. The dimensions of the EDM notches are shown in Table~\ref{tab_notchSize}. Sub-aperture defects were selected to experimentally evaluate the techniques ability to characterise defects at this limit where detection and characterisation typically breaks down. The ability for this technique to accurately characterise sub-aperture defects was therefore evaluated experimentally. 

\begin{table}[!h]
\caption{X-ray CT measured notch dimensions}
\label{tab_notchSize}
\centering\begin{tabular}{lccc}
\hline
Defect No. & Length $\pm 0.03$ mm & Width $\pm 0.03$ mm & Depth $\pm 0.03$ mm \\
\hline
1 & $1.02$ & $0.14$ & $0.45$ \\
2 & $2.01$ & $0.13$ & $0.45$ \\
3 & $2.97$ & $0.14$ & $0.47$ \\\hline
\end{tabular}
\vspace*{-4pt}
\end{table}

A parametric database of EDM notch-defect ECT scan results were simulated using the hybrid-model developed in section~\ref{sec_sim}, and PCA performed to generate manifolds in N-dimensional pc-space. The simulated notches have a fixed surface width of $0.13 mm$ while their other parameters, length ($l$) and depth ($d$), were varied to produce a parametric database of defect scans ($1.0 \leq l \leq 5.0 mm$, $0.1 \leq d \leq 1.3 mm$). For each defect a 2D scan is simulated with a spatial resolution of $\Delta x = \Delta y = 0.5\ mm$.  Due to the axial symmetry of the defect and the coil, only a single quadrant of the 2D scan is required. The full scan can then be reconstructed through reflection of the quadrant about the origin at the center of the defect.  

The parametric database of $L\times D$ defect scans was produced, where $L$ \& $D$ represent the defect length and depth parameter ranges respectively. The linearly-distributed ranges and parameter sizes of the simulated database are shown in Table~\ref{tab_2Dparams}.

\begin{table}[!h]
\caption{Simulated defect scan parametric database ranges and size}
\label{tab_2Dparams}
\centering\begin{tabular}{lccc}
\hline
Parameter & Range & Increment & Size \\
\hline
$X$ & $-4.5 \le x \le 4.5 mm$ & $\Delta x = 0.5 mm$ & 19  \\
$Y$ & $-4.5 \le y \le 4.5 mm$ & $\Delta y = 0.5 mm$ & 19  \\
$F$ & $1.0 \le f \le 8.0 MHz$ & $\Delta f = 0.08 MHz$ & 90   \\
$L$ & $1.0 \le l \le 5.0 mm$ & $\Delta l = 0.5 mm$ & 9  \\
$D$ & $1.0 \le l \le 1.3 mm$ & $\Delta d = 0.3 mm$ & 5  \\\hline
\end{tabular}
\vspace*{-4pt}
\end{table}

Applying the PCA as detailed in section~\ref{sec_pca} to the defect database at a given frequency, defect characterisation of the experimental defect scans. 

\subsubsection{Experimental Defect Characterisation}
The defects shown in Figure~\ref{fig_notch-CT} and Table~\ref{tab_notchSize} we inspected using the probe and impedance analyser configuration detailed in Section~\ref{sec_expset} over the range of frequencies shown in Figure~\ref{tab_2Dparams}.  Five repeat inspections were performed over each defect as shown in Figure~\ref{fig_notch-CT}.  For each repeat the probe was re-positioned to ensure different realisations of the same defect thus achieving realistic variations in inspection data for the same defect dimensions. This would provide more accurate determination of characterisation error of the manifold mapping approach.  Both the simulated database and the experimental data were normalised to the peak magnitude of a defect of length 3.0mm and depth 0.4 mm selected from the database and the experimental scans respectively.

\begin{figure}[!h]
\centering\includegraphics[width=6.5in]{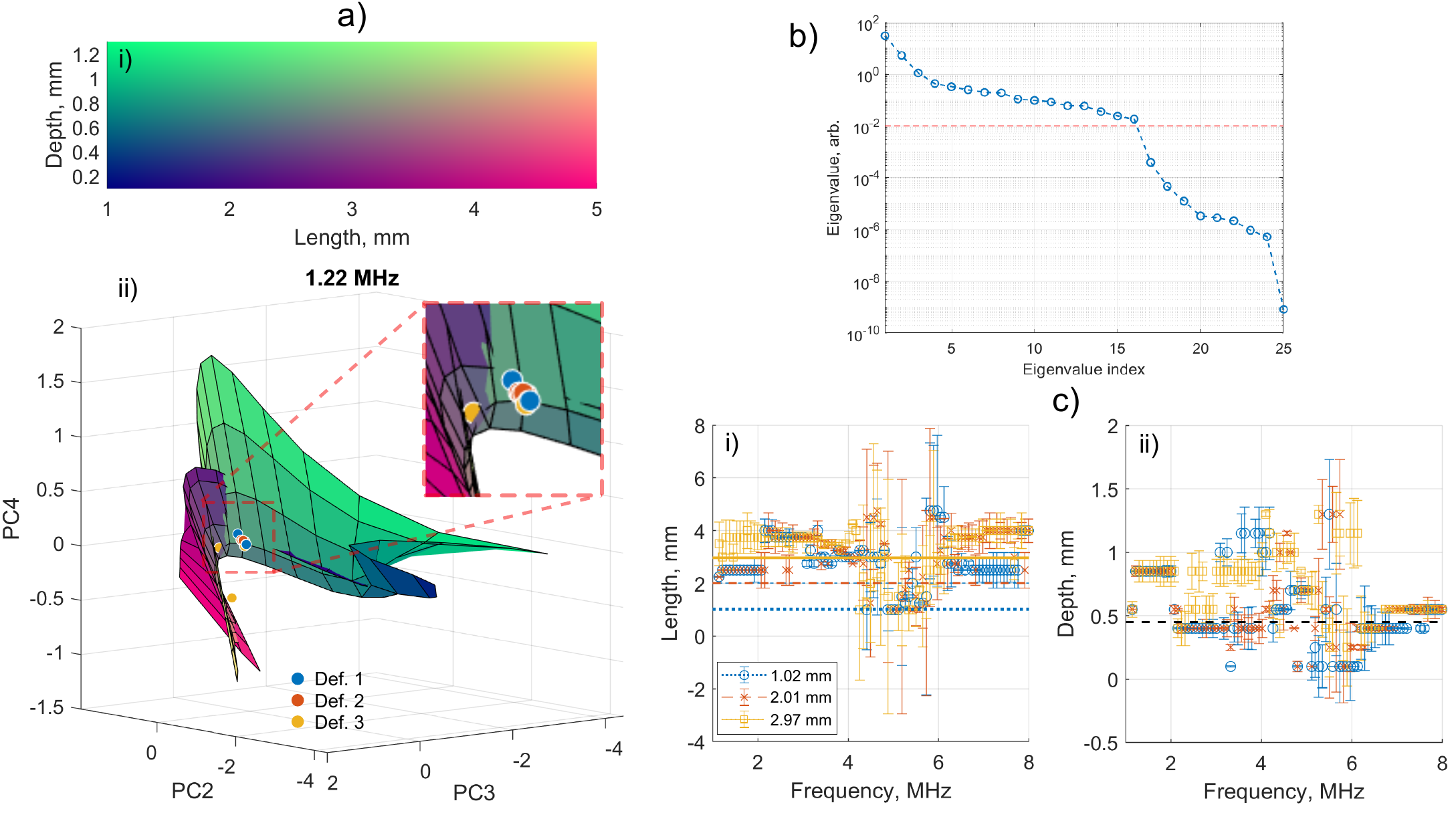}
\caption{Experimental sub-diameter defect characterisation in aluminium. a.i) shows the defect parameter space of the simulated database, which is mapped into 3D principle-component space at 1.22MHz in a.ii) which shows the experimental measurements for each experimental defect measurement relative to the defect manifold. b) shows the eigenvalues of the parametric database principle components along with the $10^{-2}$ threshold. c) shows the mean predicted notch defect length (i) \& depth (ii) as a function of excitation frequency based on 16-dimensional space Euclidean distance characterisation, along with standard deviation error bounds based on the 5 repeat experimental measurements for each notch size. Horizontal Lines denote the X-ray CT measured defect dimensions.}
\label{fig_notchResults}
\end{figure}

Figure~\ref{fig_notchResults} summarises the results of notch characterisation between $1-8 MHz$, using the imaginary component of the eddy-current measurement.  Figure~\ref{fig_notchResults}.a.ii shows the parametric manifold in 3D pc-space at $1.22 MHz$ with repeated experimental defect measurements shown relative to the manifold, while Figure~\ref{fig_notchResults}.a.i represents the colourmap of the parametric database. Figure~\ref{fig_notchResults}.a.ii shows how the manifold is contorted in specific parts of the parameter space, making accurate characterisation more difficult. The close distribution of experimental measurements in 3D pc-space further demonstrates the challenges in characterising sub-aperture defects. 
Figure~\ref{fig_notchResults}.b shows the pc eigenvalues relative to a 0.01 threshold for the simulated database of defects, demonstrating that there are 16 pc's above this threshold, resulting in a 16-dimensional pc-manifold in which to characterise the experimental defect measurements. Figure~\ref{fig_notchResults}.c shows the mean predicted defect length (i) and depth (ii) of the experimental defect scans as a function of excitation frequency, along with the standard deviation characterisation error across the 5 repeat experimental scans per defect. These predicted values were determined by Euclidian distance in 16-dimensional pc-space of experimental points to the nearest point on the defect manifold.  

The are 2 clear observations that can be made about this approach to defect characterisation. The first is that sub-aperture defects are indistinguishable using this characterisation approach due to the distribution of experimental measurements in pc-space along with the complexity of the manifold surface in pc-space.  As a result, characterisation at sub-resonant frequencies determines the defects as having approximately the same length as the diameter of the coil, between $2.0 - 4.0 mm$, while depth estimation varies much more significantly within the parameter space, between $0.4 - 1.3 mm$.  Examination of the full frequency range offers no insights into which frequencies might offer greater characterisation accuracy, but does confirm that frequencies around resonance exhibit significant instabilities that cannot be modelled accurately in the simulation, making characterisation within this range near-impossible.

\section{Conclusions}
A hybrid finite-element \& equivalent circuit model for inductive eddy-current measurements across a range of MHz frequencies has been developed and validated experimentally.  The process demonstrates the approach for calibrating unknown parameters in the probe circuit in order to simulate eddy-current accurate scans over a range of common frequencies for surface breaking defect detection.

This model was used to explore a parametric manifold mapping defect characterisation approach using databases of simulated notch scans.  This approach is successfully applied to eddy-current scan data and slot depth characterisation in Aluminium, demonstrating a superior characterisation accuracy compared to standard amplitude inversion methods, particularly for shallow slot depths.  The limits of this approach were highlighted by defining the maximum depth characterisation possible on a sloped slot. Here it was shown that accurate depth characterisation is possible up to 2.0 mm in slot depth.  This depth is significantly higher than the standard depth of penetration of the measurement in Aluminium but corresponds approximately to the geometric dimensions of the inspection coil.

The model was further tested on notches of finite length, smaller than the outer diameter of the inspection coil, to explore the limits of the characterisation capability. The parametric-manifold mapping method was used to characterise the defect dimensions by reducing the dimensionality of 2D scans down to 16 principle components.  The results highlight the inherent difficulties in characterising sub coil-diameter defects demonstrated by the complexity of the parametric manifold of the simulated database in principle-component space, and through the tight distribution of experimental measurements of different defect dimensions. Although the defect characterisation method and the parameter-space sampling was not optimised for this ECT application, the limiting factor remains the poor separation between sub-aperture parameters in lower-dimensional space which is an fundamental feature dependent on the coil dimensions that will be difficult to overcome.

\section{Acknowledgements}
The authors would like to thank Prof. John Bowler \& Dr Alexander Velichko for their advise and insight, as well as James Wilcox from Rolls-Royce plc (NDT lab) for sizing and providing X-ray CT images of test specimen defects.  This work was funded via the Research Centre for Non-destructive Evaluation (grant number EP/L022125/1). The data presented in this paper can be found at the University of Bristol Research Data Storage Facility (RDSF) - DOI (PROVIDED UPON ACCEPTANCE).

\vskip6pt

\enlargethispage{20pt}


\bibliographystyle{ieeetr}
\bibliography{Master.bib}

\begin{thebibliography}{10}

\bibitem{Bray1992}
D.~Bray and D.~Mcbride, {\em Nondestructive testing techniques}.
\newblock New York, NY (United States); John Wiley Sons, Inc., Jan 1992.

\bibitem{Sabbagh2013}
H.~A. Sabbagh, R.~K. Murphy, E.~H. Sabbagh, J.~C. Aldrin, and J.~S. Knopp,
  ``{Overview of Methods of Computational Electromagnetics},'' pp.~3--5, 2013.

\bibitem{Ahmed2017}
S.~Ahmed, M.~Salucci, R.~Miorelli, N.~Anselmi, G.~Oliveri, P.~Calmon,
  C.~Reboud, and A.~Massa, ``{Real time groove characterization combining
  partial least squares and SVR strategies: Application to eddy current
  testing},'' in {\em Journal of Physics: Conference Series}, 2017.

\bibitem{Aldrin2014}
J.~C. Aldrin, C.~Annis, H.~A. Sabbagh, J.~S. Knopp, and E.~A. Lindgren,
  ``{Assessing the reliability of nondestructive evaluation methods for damage
  characterization},'' in {\em AIP Conference Proceedings}, 2014.

\bibitem{Cherry2014}
M.~R. Cherry, S.~Sathish, A.~L. Pilchak, A.~J. Cherry, and M.~P. Blodgett,
  ``Characterization of microstructure with low frequency electromagnetic
  techniques,'' {\em AIP Conference Proceedings}, vol.~1581, no.~1,
  pp.~1456--1462, 2014.

\bibitem{Salucci2017}
M.~Salucci, L.~Poli, and A.~Massa, ``{Advanced multi-frequency GPR data
  processing for non-linear deterministic imaging},'' {\em Signal Processing},
  2017.

\bibitem{Rocca2009}
P.~Rocca, M.~Benedetti, M.~Donelli, D.~Franceschini, and A.~Massa,
  ``{Evolutionary optimization as applied to inverse scattering problems},''
  2009.

\bibitem{Ahmed2019}
S.~Ahmed, C.~Reboud, P.~E. Lhuillier, P.~Calmon, and R.~Miorelli, ``{An
  adaptive sampling strategy for quasi real time crack characterization on eddy
  current testing signals},'' {\em NDT and E International}, 2019.

\bibitem{Miorelli2019}
R.~Miorelli, A.~Skarlatos, and C.~Reboud, ``{A machine learning approach for
  classification tasks of ECT signals in steam generator tubes nearby support
  plate},'' in {\em AIP Conference Proceedings}, vol.~2102, p.~90004, American
  Institute of Physics Inc., may 2019.

\bibitem{Rosado2013}
L.~S. Rosado, F.~M. Janeiro, P.~M. Ramos, and M.~Piedade, ``{Defect
  characterization with eddy current testing using nonlinear-regression feature
  extraction and artificial neural networks},'' {\em IEEE Transactions on
  Instrumentation and Measurement}, 2013.

\bibitem{Buck2016}
J.~A. Buck, P.~R. Underhill, J.~E. Morelli, and T.~W. Krause, ``{Simultaneous
  Multiparameter Measurement in Pulsed Eddy Current Steam Generator Data Using
  Artificial Neural Networks},'' {\em IEEE Transactions on Instrumentation and
  Measurement}, 2016.

\bibitem{Shokralla2016}
S.~Shokralla, J.~E. Morelli, and T.~W. Krause, ``{Principal Components Analysis
  of Multifrequency Eddy Current Data Used to Measure Pressure Tube to
  Calandria Tube Gap},'' {\em IEEE Sensors Journal}, 2016.

\bibitem{DAngelo2018}
G.~D'Angelo, M.~Laracca, S.~Rampone, and G.~Betta, ``{Fast Eddy Current Testing
  Defect Classification Using Lissajous Figures},'' {\em IEEE Transactions on
  Instrumentation and Measurement}, 2018.

\bibitem{Cai2018}
C.~Cai, R.~Miorelli, M.~Lambert, T.~Rodet, D.~Lesselier, and P.~E. Lhuillier,
  ``{Metamodel-based Markov-Chain-Monte-Carlo parameter inversion applied in
  eddy current flaw characterization},'' {\em NDT and E International}, 2018.

\bibitem{Douvenot2011}
R.~Douvenot, M.~Lambert, and D.~Lesselier, ``{Adaptive metamodels for crack
  characterization in eddy-current testing},'' {\em IEEE Transactions on
  Magnetics}, 2011.

\bibitem{Salucci2019}
M.~Salucci, N.~Anselmi, G.~Oliveri, P.~Rocca, S.~Ahmed, P.~Calmon, R.~Miorelli,
  C.~Reboud, and A.~Massa, ``{A nonlinear Kernel-based adaptive
  learning-by-examples method for robust NDT/NDE of conductive tubes},'' {\em
  Journal of Electromagnetic Waves and Applications}, 2019.

\bibitem{Velichko2017}
A.~Velichko, L.~Bai, and B.~W. Drinkwater, ``{Ultrasonic defect
  characterization using parametric-manifold mapping},'' {\em Proceedings of
  the Royal Society A: Mathematical, Physical and Engineering Science}, 2017.

\bibitem{MarkusBuehren}
{Markus Buehren}, ``{Differential Evolution - File Exchange - MATLAB
  Central}.''

\bibitem{Smith2004}
R.~A. Smith and D.~J. Harrison, ``Hall sensor arrays for rapid large-area
  transient eddy current inspection,'' {\em Insight - Non-Destructive Testing
  and Condition Monitoring}, vol.~46, no.~3, pp.~142--146, 2004.

\end{thebibliography}

\end{document}